\documentclass[12pt]{iopart}
\usepackage{graphicx}

\begin{document}

\title{Non-Hermitian quantum mechanics: the case of bound state scattering
theory}
\author{A.\ Matzkin }

\address{Laboratoire de Spectrom\'{e}trie physique (CNRS
Unit\'{e} 5588), Universit\'{e} Joseph-Fourier Grenoble-1, BP 87,
38402 Saint-Martin, France}

\begin{abstract}
Excited bound states are often understood within scattering based
theories as resulting from the collision of a particle on a target
via a short-range potential. We show that the resulting formalism
is non-Hermitian and describe the Hilbert spaces and metric
operator relevant to a correct formulation of such theories. The
structure and tools employed are the same that have been
introduced in current works dealing with PT-symmetric and
quasi-Hermitian problems. The relevance of the non-Hermitian
formulation to practical computations is assessed by introducing a
non-Hermiticity index. We give a numerical  example involving
 scattering by a short-range potential in a Coulomb field for which it is seen
that even for a small but non-negligible non-Hermiticity index the
non-Hermitian character of the problem must be taken into account.
The computation of physical quantities in the relevant Hilbert
spaces is also discussed.

\end{abstract}


\maketitle

\section{Introduction}

The standard formulation of quantum mechanics requires physical
observables to be mathematically given in terms of Hermitian
operators. In recent years theories with a non-Hermitian Hamiltonian
have been receiving an increasing interest sparked by work in the
field of PT-symmetric quantum mechanics \cite{bender05}.
PT-symmetric Hamiltonians are complex but nevertheless possess a
real spectrum.\ The structure of PT-symmetric theories, initially
suggested to hinge on the existence of a charge conjugation operator
\cite{bender etal04} has been clarified by showing \cite{mosta03}
that the non-Hermitian Hamiltonians could be mapped to Hermitian
ones, and therefore be fitted within the better-known framework of
quasi-Hermitian operators \cite{scholtz etal92}. The relevance of
the non-Hermitian formulation for the description of physical
systems is still being debated \cite{mostafazadeh
batal04,jones05,mosta05,bender etal06}.

In the present work we show that the effective Hamiltonians
appearing in certain theories dealing with bound state scattering by
a short-range potential are non-Hermitian. In this case the
Hamiltonians are real and their non-Hermitian character stems from
the boundary conditions obeyed by the eigenstates: on the one hand
there is no physical asymptotic freedom (since the states are bound)
and on the other the scattering solutions inside the short-range
potential region do not exist. However contrarily to the situation
in PT-symmetric problems, there is in principle an underlying
Hermitian Hamiltonian, whose solutions are unknown in practice but
whose existence may provide guiding rules when undertaking practical
computations. We will see that the unambiguous formulation of bound
state scattering sheds some light on issues regarding the physical
relevance of non-Hermitian formulations of quantum mechanics. Let us
mention that in the overwhelming majority of applications of the
bound state scattering formalism to nuclear, atomic or molecular
physics non-Hermitian issues have been downright ignored; this is
unproblematic when non-Hermiticity is small (as is generally the
case), but we will give an illustration in which ignoring the
non-Hermitian nature of the scattering  Hamiltonian brings in errors
that can be directly attributed to the (inappropriate) use of the
standard inner product.

We will first introduce bound state scattering theory and show why
the scattering Hamiltonian is non-Hermitian in the 'physical'
Hilbert space  $\mathcal{H}_{ph}$ (Sec 2).\ The quasi-Hermitian
Hamiltonian will then be described by an expansion in terms of a
biorthogonal basis, leading naturally to the definition of a new
inner product and its associated Hilbert space $\mathcal{H}$ (Sec\
3). In line with previous works on quasi-Hemitian operators, we will
examine the relationship between the two Hilbert spaces
$\mathcal{H}_{ph}$  and $\mathcal{H}$ in terms of the metric
operator and further discuss the computation of physical results in
$\mathcal{H}$ and $\mathcal{H}_{ph}$. In Sec 4, the formalism will
be illustrated by carrying out the numerical calculation of an
experimentally observable quantity (the autocorrelation function) in
the particular case of short-range scattering in a Coulomb field.
Our concluding remarks will be given in Sec\ 5.

\section{Scattering description of excited bound states}

Let $H^{e}$ be the exact Hamiltonian of the 2-particle scattering problem
(in the center of mass; the physical situation most often considered is that
of a light particle colliding on a massive compound target). We assume $%
H^{e} $ can be split as%
\begin{equation}
H^{e}=H_{0}+V  \label{1}
\end{equation}%
where $V$ contains all the \emph{short range} interactions between
the particles.\ We further assume $H_{0}$ is spherically symmetric
(in terms of the relative coordinate) and that  short-range
 means that%
\begin{equation}
V=\theta (r_{0}-r^{\prime })V\theta (r_{0}-r),  \label{3}
\end{equation}%
ie $V$ vanishes outside some small radius $r_{0}$ ($\theta $ is the step
function). Therefore $H_{0}$ contains not only the kinetic and internal
terms of the non-interacting particles, but also any long-range interaction
between them. Let $E$ be the total energy; allowing for inelastic scattering
$E$ is partitioned as%
\begin{equation}
E=\varepsilon _{i}+\epsilon _{i}
\end{equation}%
where $\varepsilon _{i}$ and $\epsilon _{i}$ are the internal and
the kinetic energy respectively (in the case of a massive target
$\varepsilon _{i}$ depends on the internal states of the target
whereas $\epsilon _{i}$ is the collision energy of the light
particle). The eigenstates of $H_{0}$ are given by
\begin{equation}
\left| \phi _{i}(E)\right\rangle =\left| f_{i}(\epsilon _{i})\right\rangle
\left| i(\varepsilon _{i})\right\rangle ;  \label{4}
\end{equation}%
$f_{i}(\epsilon _{i},r)\equiv \left\langle r\right. \left| f_{i}(\epsilon
_{i})\right\rangle $ is the eigenfunction of the radial part of $H_{0}$
whereas the 'target' state $\left| i(\varepsilon _{i})\right\rangle $
includes all the other degrees of freedom, including the non-radial ones of
the colliding particle (a handy notation given that the angular momenta of
the particles are usually coupled). The target states are orthogonal, $%
\left\langle i\right| \left. j\right\rangle =\delta _{ij}$. Since we are
dealing with bound states, $f_{i}(\epsilon _{i},r)$ vanishes at $0$ and $%
+\infty $ (whenever $E$ is an eigenvalue of $H_{0}$).

The label $i$ defines the scattering channel. In each channel the
standing-wave solutions are given by the Lippmann-Schwinger
equations of scattering theory as
\begin{equation}
\left| \psi _{i}^{e}(E)\right\rangle =\left| \phi _{i}(E)\right\rangle
+G_{0}(E)K(E)\left| \phi _{i}(E)\right\rangle  \label{z50}
\end{equation}%
where $G_{0}(E)$ is the principal-value Green's function and $K$ the
reaction (scattering) operator for standing waves linked to the
familiar $S$ matrix by a Cayley transform \cite{newton82}. The
difference here with standard scattering theory is that the bound
channels are included explicitly \cite{fano78,jalabert,matzkin99}.\
The consequences are that (\emph{i}) $G_{0}(E)$ has no poles -- it
is modified \cite{fano78} relative to the usual resolvent by
including a term
(solution of the homogeneous equation) that has poles at the eigenvalues of $%
H^{e}$ so that overall $G_{0}(E)$ has no poles (but diverges radially) \footnote{%
As stressed by Fano \cite{fano78} who introduced this 'smooth
Green's function', for genuine scattering states (continuum
energies), $G_{0}(E)$ becomes the standard Green's function.};
(\emph{ii}) there is no asymptotic freedom:
both $\left\langle r\right| \left. \phi _{i}(E)\right\rangle $ and $%
\left\langle r\right| \left. \psi _{i}^{e}(E)\right\rangle $ diverge at $%
r\rightarrow \infty $ for an arbitrary value of $E$; (\emph{iii}) an
eigenstate of $H^{e}$ cannot be given by a single channel solution
of the form (\ref{z50})
but requires a superposition%
\begin{equation}
\left| \psi ^{e}(E)\right\rangle =\sum_{i}Z_{i}(E)\left| \psi
_{i}^{e}(E)\right\rangle   \label{z52}
\end{equation}%
where the expansion coefficients $Z_{i}(E)$ are determined by the
asymptotic
($r\rightarrow \infty $) boundary conditions such that at the eigenvalues $%
\left\langle r\right| \left. \psi ^{e}(E)\right\rangle $ vanishes at
infinity.

Formally $H^{e}\left| \psi ^{e}(E)\right\rangle =E\left| \psi
^{e}(E)\right\rangle $ is satisfied as well as the usual properties
for
eigenstates of Hermitian operators, such as their orthonormality%
\begin{equation}
\left\langle \psi ^{e}(E_{1})\right| \left. \psi
^{e}(E_{2})\right\rangle =\delta _{E_{1}E_{2}}
\end{equation}%
or the spectral decomposition theorem. However in practice the
expansion of $G_{0}$ over the eigenstates of $H_{0}$ is intractable.
Instead the radial part of $G_{0}$ is separated and the expansion
over the energies reduced to the closed form $f_{i}(\epsilon
_{i},r_{<})g_{i}(\epsilon _{i},r_{>})$; $g_{i}$ is a solution of the
radial part of $H_{0}$ irregular at the origin (for arbitrary bound
energies, both $f$ and $g$ exponentially diverge in the limit
$r\rightarrow \infty$).\ Hence the closed form of the radial Green's
function only makes sense for $r>r_{0}$ (where $V$ vanishes). This
is of course consistent with the scattering point of view: when
$r<r_{0}$ we are inside the reaction zone and there is no scattering
solution, whatever happens within the reaction zone being encoded in
the phase-shifts. The
wavefunction (\ref{z52}) outside the reaction zone becomes%
\begin{equation}
\left\langle r\right. \left| \psi (E)\right\rangle =\sum_{i}Z_{i}(E)\left[
f_{i}(\epsilon _{i},r)\left| i\right\rangle +\sum_{j}g_{j}(\epsilon
_{j},r)\left| j\right\rangle K_{ji}\right] \qquad r>r_{0}  \label{e20}
\end{equation}%
where $K_{ji}$ are the on-shell elements of the scattering matrix, which are
assumed to be known.

It is important to note that the scattering eigenstate (\ref{e20}) is the
part for $r>r_{0}$ of the exact solution $\left| \psi ^{e}(E)\right\rangle $%
, and not an approximation to it. But within the scattering fomulation the
'inner' part of $\left| \psi ^{e}(E)\right\rangle $ for $r<r_{0}$ does not
exist: all meaningful quantities are defined radially on $[r_{0},\infty
\lbrack $. Indeed let us write%
\begin{equation}
\left| \psi ^{e}(E)\right\rangle =\theta (r-r_{0})\left| \psi
(E)\right\rangle +\theta (r_{0}-r)\left| \psi _{inner}(E)\right\rangle
\label{a20}
\end{equation}%
and let
\begin{equation}
H\equiv \sum_{E}E\left| \psi (E)\right\rangle \left\langle \psi (E)\right|
\label{a25}
\end{equation}%
be the restriction of $H^{e}$ to the outer region $r>r_{0}$. $H$
 is the only operator directly known from the solutions of the scattering problem. We
have the following
properties :%
\begin{eqnarray}
H=H^{+}  \label{e23a} \\
\left\langle \psi (E_{1})\right| \left. \psi (E_{2})\right\rangle =\delta
_{E_{1}E_{2}}+\mu _{E_{1}E_{2}}(1-\delta _{E_{1}E_{2}})  \label{e23b} \\
H\left| \psi (E)\right\rangle \neq E\left| \psi (E)\right\rangle .
\label{e23c}
\end{eqnarray}%
That $H$ is Hermitian relative to the standard product can be seen to follow
from its definition (\ref{a25}). Eq. (\ref{e23b}) tells us first that the $%
\left| \psi (E)\right\rangle $ are normalized to 1 like the $\left| \psi
^{e}(E)\right\rangle $ which might appear surprising in view of (\ref{a20})
but follows by showing normalization does not depend on the inner radial
part of the wavefunction (this is done by expressing the normalization
integral in terms of radial Wronskians, see Sec 5.7 of \cite{fano}). $\ $Eq (%
\ref{e23b}) also indicates that the scattering eigenstates $\left| \psi
(E)\right\rangle $ are not orthogonal since the scalar product of two
disitinct eigenstates is given by $\mu $. This may be shown by rearranging
eq (\ref{e20}) in the form%
\begin{equation}
\left\langle r\right. \left| \psi (E)\right\rangle =\sum_{i}X_{i}(E)\left|
i(\varepsilon _{i})\right\rangle F_{i}(\epsilon _{i},r)\qquad r>r_{0},
\label{e24}
\end{equation}%
where the overall contribution in a given scattering channel $i$ is
grouped together. As a consequence the radial channel functions
$F_{i}(\epsilon _{i},r)$ must vanish as $r\rightarrow \infty$ for
each $i$ (the scattering information is now contained in the $F$
functions and in the new coefficients $X$ that both depend on $K$).\
Recalling the target states are orthogonal, the scalar product
(\ref{e23b}) is seen to depend solely on the radial
overlaps between identical channel radial functions at different energies, given by%
\begin{equation}
\left\langle F_{i}(\epsilon _{1})\right| \left. F_{i}(\epsilon _{2})\right\rangle =\frac{W[F_{i}(\epsilon _{2}),F_{i}(\epsilon _{1})]_{r_{0}}}{%
\epsilon _{2}-\epsilon _{1}},  \label{e25}
\end{equation}%
where $W$ is the Wronskian taken at $r_{0}$. This equality follows
from computing $\left\langle F_{i}(\epsilon _{1})\right|
p_{r}^{2}\left| F_{i}(\epsilon _{2})\right\rangle -\left\langle
F_{i}(\epsilon _{2})\right| p_{r}^{2}\left| F_{i}(\epsilon
_{1})\right\rangle $ (integrate by parts and recall that the scalar
product is defined in $[r_{0},\infty ]$). This gives rise to nonzero
boundary terms at $r_{0}$, impliying that $p_{r}^{2}$ is not
Hermitian on $[r_{0},\infty ]$ \footnote{%
The non-Hermitian character of $\frac{d^{2}}{dr^{2}}$ on bounded
intervals with arbitrary boundary conditions is of course trivial.
In the context of scattering theory, this fact was pointed out in
particular by Bloch \cite{bloch} who introduced a singular surface
operator to cancel the boundary terms when defining quantities on
$[0,r_{0}]$.\ However the non-Hermitian character of the scattering
eigenstates on $[r_{0},+\infty ]$ is irrelevant in standard
scattering theory because the solutions of $H^{e}$ and $H_{0}$ are
both (improperly) normalized by the same asymptotic condition,
hinging on the isometry of the wave operators.}.

Because the  $\left| \psi (E)\right\rangle $ are not orthogonal,
they cannot be eigenstates of the Hermitian operator $H$ [eq
(\ref{e23c})] but are eigenvectors of a non-Hermitian Hamiltonian
denoted
 $\widetilde{H}$. From eqs (\ref{1}) and (\ref{3}) we see that $%
\widetilde{H}$ is formally given by $H_{0}$ redefined
by restricting it radially to the interval $%
[r_{0},\infty ]$ and supplementing it by specific boundary
conditions on the surface $r=r_{0}$. It is precisely this fact,
combined with the lack of asymptotic completeness, that leads to
non-Hermiticity. This completes our brief discussion on the
non-Hermitian character of the bound state scattering problem; we
now analyze the structure of the non-Hermitian theory and further
examine the implications of this non-Hermiticity in practical
problems.

\section{Quasi-Hermitian operators: metric and Hilbert spaces}

Here we forget about the existence of an underlying exact
Hamiltonian and we take the practical scattering viewpoint: the
phase-shifts are given numbers (obtained from a symmetric $K$
matrix) and the physical states are represented by vectors in
$\mathcal{H}_{ph}$, which is essentially the Hilbert space of
standard quantum mechanics: it is endowed with the standard scalar
product except that radially the integral is defined on
$[r_{0},\infty ]$. This slight modification of the radial integral
does not cause any difference since the states of interest in
scattering phenomena (such as Gaussian states) have
negligible probability amplitude in the inner zone. In this sense the $%
\left| \psi (E)\right\rangle $ belong to\ $\mathcal{H}_{ph}$.

Since $%
\widetilde{H}$ is non-Hermitian on $\mathcal{H}_{ph}$, we have%
\begin{eqnarray}
\left\langle \psi (E^{\prime })\right| \widetilde{H}\left| \psi
(E)\right\rangle &=&E\left\langle \psi (E^{\prime })\right. \left| \psi
(E)\right\rangle \label{eep}\\
\left\langle \psi (E^{\prime })\right| \widetilde{H}^{+}\left| \psi
(E)\right\rangle &=&E^{\prime }\left\langle \psi (E^{\prime })\right. \left|
\psi (E)\right\rangle .
\end{eqnarray}%
We are thus naturally lead to introduce a biorthogonal set $\{\left|
\widetilde{\psi }(E)\right\rangle ,\left| \psi (E)\right\rangle \}$
\cite{morse}, where we denote by $\left| \widetilde{\psi
}(E)\right\rangle $ the
eigenstates of $\widetilde{H}^{+}$. The following properties are satisfied:%
\begin{eqnarray}
&&\widetilde{H}\left| \psi (E)\right\rangle =E\left| \psi (E)\right\rangle
\label{e31} \\
&&\widetilde{H}^{+}\left| \widetilde{\psi }(E)\right\rangle =E\left|
\widetilde{\psi }(E)\right\rangle  \label{e32} \\
&&\left\langle \widetilde{\psi }(E)\right. \left| \psi (E^{\prime
})\right\rangle =\delta _{EE^{\prime }}  \label{e33}
\end{eqnarray}%
from which it follows that we can write the following expansions:%
\begin{equation}
\widetilde{H}=\sum_{E}E\left| \psi (E)\right\rangle \left\langle \widetilde{%
\psi }(E)\right| \qquad \widetilde{H}^{+}=\sum_{E}E\left| \widetilde{\psi }%
(E)\right\rangle \left\langle \psi (E)\right| .  \label{e34}
\end{equation}%
$\widetilde{H}$ and $\widetilde{H}^{+}$ are further linked by%
\begin{equation}
\widetilde{H}=\mathcal{G}\widetilde{H}^{+}\mathcal{G}^{-1}  \label{e40}
\end{equation}%
where $\mathcal{G}$ is a Hermitian operator given by%
\begin{eqnarray}
\mathcal{G} &=&\sum_{E}\left| \psi (E)\right\rangle \left\langle \psi
(E)\right|  \label{e41} \\
\mathcal{G}^{-1} &=&\sum_{E}\left| \widetilde{\psi }(E)\right\rangle
\left\langle \widetilde{\psi }(E)\right| .  \label{e42}
\end{eqnarray}
We will take for granted the completeness of the biorthogonal basis,
although it is by no means obvious.\ In particular the difficulties
that arise when $\mathcal{H}_{ph}$ is of infinite dimensions have
been pointed out recently \cite{ks04,tanaka06}. Completeness of the
biorthogonal basis implies that the 'canonical metric basis' (in the
sense of \cite{mostafazadeh batal04}), consisting of the
eigenvectors of the metric operator, is also complete. From there we
deduce that an arbitrary state of $\mathcal{H}_{ph}$ can in
principle be expanded in terms of the $\left| \psi (E)\right\rangle
$, ie the eigenstates of $\widetilde{H}$ span the entire Hilbert
space of admissible physical states even if they do not form an
orthogonal basis in $\mathcal{H}_{ph}$.

The relations (\ref{eep})--(\ref{e42}) have become familiar lately
in the context of PT-symmetric quantum mechanics and more largely in
works dealing with quasi-Hermitian operators (see in particular
\cite{mostafazadeh batal04}). Eq (\ref{e40}) is the defining
relation of quasi-Hermiticity \cite{ks04} provided $\mathcal{G}$ is
invertible ($\mathcal{G}^{-1}$ then being its inverse, since by (\ref{e33}) $%
\mathcal{GG}^{-1}$ is a representation of the unit operator in $\mathcal{H}$%
) and positive-definite. We will not attempt to prove these properties
here.\ We note however that if the $\left| \psi (E)\right\rangle $ (and
hence the $\left| \widetilde{\psi }(E)\right\rangle $) form a basis of $%
\mathcal{H}_{ph}$, as we have assumed to be the case, then
$\mathcal{G}$ has no null eigenvalue and is thus invertible. It is
of course a working hypothesis in scattering theory that any
meaningful physical state can be expanded in terms of the $\left|
\psi (E)\right\rangle $ (but this may not be true mathematically for
a given arbitrary vector).\ The positive-definiteness of
$\mathcal{G}$ follows heuristically by remarking that in the 'mixed'
representation
\begin{equation}
\left\langle \widetilde{\psi }(E^{\prime })\right| \mathcal{G}\left| \psi
(E)\right\rangle  \label{e43}
\end{equation}%
simply becomes (\ref{e23b}), so that $\mathcal{G}\geq I+\mu M$
where $I$ is the identity matrix, $M$ is the special matrix with elements $%
M_{ij}=1-\delta _{ij}$ and $\mu $ a small ($|\mu |\ll 1$) real
number. The positive-definiteness of $I+\mu M$ ensures that
$\mathcal{G}$ is positive definite too. The positive-definitiness of
$\mathcal{G}$ is important to define a positive norm in
$\mathcal{H}$ \cite{scholtz etal92,mostafazadeh batal04,ks04}. Since
from eq (\ref{e42})
\begin{equation}
\left| \widetilde{\psi }(E)\right\rangle =\mathcal{G}^{-1}\left| \psi
(E)\right\rangle ,  \label{e44}
\end{equation}%
the inner product is defined through
\begin{equation}
\left( \psi (E_{1}),\psi (E_{2})\right) _{\mathcal{G}}\equiv \left\langle
\psi (E_{1})\right| \mathcal{G}^{-1}\left| \psi (E_{2})\right\rangle
=\left\langle \widetilde{\psi }(E_{1})\right. \left| \psi
(E_{2})\right\rangle =\delta _{E_{1}E_{2}}.  \label{e45}
\end{equation}
$\mathcal{G}$ is thus seen to be (the positive definite) metric. By eq (\ref%
{e40}) it is immediate to verify that $\widetilde{H}$ is Hermitian
relative to this new inner product.

Let $\mathcal{H}$ be the Hilbert space endowed with the inner
product defined by (\ref{e45}). Calculations are simple to perform
in $\mathcal{H}$
because the new scalar product reestablishes orthogonality. Indeed \ let $%
\left| \phi _{1}\right\rangle =\sum \alpha _{1}(E)\left| \psi
(E)\right\rangle $ and $\left| \phi _{2}\right\rangle =\sum \alpha
_{2}(E)\left| \psi (E)\right\rangle $ be two vectors in $\mathcal{H}$. Then
it follows from eq (\ref{e45}) that%
\begin{eqnarray}
\left( \phi _{1},\phi _{1}\right) _{\mathcal{G}} &=&\left\langle \widetilde{%
\phi }_{1}\right. \left| \phi _{1}\right\rangle =\sum_{E}\left| \alpha
_{1}(E)\right| ^{2}=1 \\
\left( \phi _{1},\phi _{2}\right) _{\mathcal{G}} &=&\left\langle \widetilde{%
\phi }_{1}\right. \left| \phi _{2}\right\rangle =\sum_{E}\alpha _{1}^{\ast
}(E)\alpha _{2}(E)
\end{eqnarray}%
with the obvious notation%
\begin{equation}
\left| \widetilde{\phi }_{1}\right\rangle \equiv \mathcal{G}^{-1}\left| \phi
_{1}\right\rangle =\sum_{E}\alpha _{1}(E)\mathcal{G}^{-1}\left| \psi
(E)\right\rangle .
\end{equation}%
We further see that quantities involving the expansions of the
non-Hermitian Hamiltonian, such as the time evolution operator,
cannot be directly determined in $\mathcal{H}_{ph}$, precisely
because of the non-Hermiticity of $\widetilde{H}$ in
$\mathcal{H}_{ph}$. But in $\mathcal{H}$ the evolution
operator is given by%
\begin{equation}
\widetilde{U}(t)=\sum_{E}e^{-iEt}\left| \psi (E)\right\rangle \left\langle
\widetilde{\psi }(E)\right| .  \label{e50}
\end{equation}%
Hence for example if we take an initial state as $\left| \phi
(t=0)\right\rangle =\left| \phi _{1}\right\rangle $, the state evolves
according to
\begin{equation}
\left| \phi (t)\right\rangle =\sum_{E}e^{-iEt}\left| \psi (E)\right\rangle
\left( \psi (E),\phi _{1}\right) _{\mathcal{G}}=\sum_{E}e^{-iEt}\alpha
_{1}(E)\left| \psi (E)\right\rangle ,
\end{equation}%
operating in effect in $\mathcal{H}$ as we would in $\mathcal{H}_{ph}$ with
a Hermitian operator.

However, in scattering problems, the physical states are known in
$\mathcal{H}_{ph}$, not in $\mathcal{H}$. Let $\left| \zeta
_{1}\right\rangle $ and $\left| \zeta _{2}\right\rangle $ be two vectors in $%
\mathcal{H}_{ph}$ and assume they can be expanded over the $\left| \psi
(E)\right\rangle $ as $\left| \zeta _{i}\right\rangle =\sum a_{i}(E)\left|
\psi (E)\right\rangle $. They are normalized relative to the standard scalar
product,%
\begin{equation}
\left\langle \zeta _{i}\right. \left| \zeta _{i}\right\rangle
=1=\sum_{EE^{\prime }}a_{i}^{\ast }(E)a_{i}(E^{\prime })\left\langle \psi
(E)\right. \left| \psi (E^{\prime })\right\rangle ;
\end{equation}%
since the basis is nonorthogonal in $\mathcal{H}_{ph},$ $\sum_{E}\left|
a_{i}(E)\right| ^{2}\neq 1$. On the other hand operators involving the
Hamiltonian, such as the evolution operator (\ref{e50}) are known on $%
\mathcal{H}$ but not on $\mathcal{H}_{ph}$. The transformation between the
two Hilbert spaces can be done both ways, $\mathcal{H}\longrightarrow
\mathcal{H}_{ph}$ for the operators or $\mathcal{H}_{ph}\longrightarrow
\mathcal{H}$ for the states.\ Indeed if an operator $\widetilde{A}$ is
Hermitian in $\mathcal{H}$ then
\begin{equation}
A=\mathcal{G}^{-1/2}\widetilde{A}\mathcal{G}^{1/2}  \label{e53}
\end{equation}%
is Hermitian in $\mathcal{H}_{ph}$. This follows directly from the general
version of eq (\ref{e40}),%
\begin{equation}
\widetilde{A}=\mathcal{G}\widetilde{A}^{+}\mathcal{G}^{-1}.
\end{equation}%
This transformation defines a linear map \cite{mostafazadeh batal04} that leaves
the inner product invariant:%
\begin{equation}
\left( \phi _{1},\phi _{2}\right) _{\mathcal{G}}=\left\langle \widetilde{%
\phi }_{1}\right. \left| \phi _{2}\right\rangle =\left\langle \phi
_{1}\right| \mathcal{G}^{-1}\left| \phi _{2}\right\rangle =\left\langle
\zeta _{1}\right| \left. \zeta _{2}\right\rangle  \label{e55}
\end{equation}%
where we have defined%
\begin{equation}
\left| \zeta _{i}\right\rangle \equiv \mathcal{G}^{-1/2}\left| \phi
_{i}\right\rangle .  \label{e57}
\end{equation}%
Therefore $\left| \zeta _{i}\right\rangle $ and $\left| \phi
_{i}\right\rangle $ represent the \emph{same} physical state but relative to
different Hilbert spaces: $\left| \zeta _{i}\right\rangle $ in $\mathcal{H}%
_{ph}$ and $\left| \phi _{i}\right\rangle $ in $\mathcal{H}$. Of
course as vectors we may as well have for instance $\left| \phi
_{i}\right\rangle \in \mathcal{H}_{ph}$ but then $\left| \phi
_{i}\right\rangle $ does not describe the same physical state as it
does in $\mathcal{H}$. It is interesting to note that the
 functions  $\left| \psi^{e}(E) \right\rangle$ defined on $\mathcal{H}_{ph}$
  (with $r \in [0,+\infty]$) represent the \emph{exact} eigenstates of the underlying Hamiltonian.
But the $\left|\psi(E) \right\rangle$ envisaged as the restriction
for $r>r_0$ of the  $\left| \psi^{e}(E) \right\rangle$ do not
represent the eigenstates in  $\mathcal{H}_{ph}$
  (now with $r \in [r_0,+\infty]$) but in $\mathcal{H}$, that is on the
  Hilbert space in which the Hamiltonian $\widetilde{H}$ is
Hermitian, despite the fact that $\left\langle r \right.%
\left|\psi(E)\right\rangle$ and  $\left\langle r \right.%
\left|\psi^{e}(E)\right\rangle$ are identical for $r>r_0$.

Finally, we briefly describe how to undertake practical calculations.
Recalling that scattering solutions as well as the Hamiltonian $\widetilde{H}
$ are defined in $\mathcal{H},$ and comparing eqs (\ref{e53}) and (\ref{e57}%
), it appears that it is computationally simpler to transform the
physical states from $\mathcal{H}_{ph}$ to $\mathcal{H}$ rather than
transform the operators to $\mathcal{H}_{ph}$. Nevertheless in both
cases it is necessary to determine the metric $\mathcal{G}$. In
general (as in the illustration given below) $\mathcal{G}$ is a
matrix of infinite rank=. $\mathcal{G}$ is therefore truncated
around the energy interval of interest.\ The matrix
elements are determined in the 'mixed' representation given by eq (\ref{e43}%
), which simply amounts to determine the overlaps%
\begin{equation}
\mathcal{G}_{EE^{\prime }}=\left\langle \psi (E)\right| \left. \psi
(E^{\prime })\right\rangle \label{ezz1}
\end{equation}%
where $E$ and $E^{\prime }$ span the (truncated) finite interval. The
resulting matrix $\mathcal{G}^{-1}$ is numerically inverted, allowing to
determine the second set of the biorthogonal basis by eq (\ref{e44}). $%
\mathcal{G}$ can also be diagonalized, retrieving in a single step
$\mathcal{G}^{-1}$, $\mathcal{G}^{-1/2}$ and $\mathcal{G}^{1/2}$; we
then compute the operators in $\mathcal{H}_{ph}$ or the
representation of the physical states in $\mathcal{H}$ by inverting
eq (\ref{e57}). The degree of
non-Hermiticity is assessed through the metric in the mixed representation (%
\ref{ezz1}): if the Hamiltonian is Hermitian relative to the
standard inner product, $\mathcal{G}$ becomes the identity matrix.\
As non-Hermiticity becomes important, the off-diagonal elements of
the metric increase. To assess the degree of non-Hermiticity we
introduce a non-Hermiticity index $\kappa $ that we define somewhat
arbitarily by the average of the $N$ largest absolute values of $\mathcal{G}%
-I$ \ (ie the $N$ largest off-diagonal terms of the metric) where
$N$ is the dimension of the chunk of $\mathcal{G}$ under study.
$\kappa $ is thus a local spectral measure of non-Hermiticity.

\section{Illustration}

\begin{figure}[tb]
\begin{center}
\includegraphics[height=2.1in,width=3in]{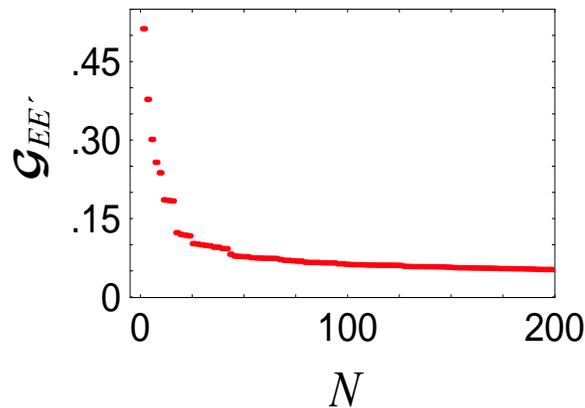}
\caption[]{Each dot represents one of the 200 ($N/2$) largest
off-diagonal elements $\left|{\cal G}_{EE^{\prime }}\right|$ of the
metric. All the diagonal elements are given by ${\cal G}_{EE}=1$.}
\label{fig1}
\end{center}
\end{figure}

To illustrate the formalism given above we will take an example in the
context of the bound states formed by the scattering of an electron on a
positively charged target.\ This situation is widely employed in atomic
physics to study the highly excited ('Rydberg') states of atoms with a
single excited electron. More specifically we will compute the
autocorrelation function
\begin{equation}
C(t)=\left\langle \zeta (t=0)\right| \left. \zeta (t)\right\rangle
\label{e63}
\end{equation}%
in two ways: by taking into account the non-Hermitian character of
the Hamiltonian on the one hand, and by downright ignoring
Hermiticity related issues on the other hand. $\left| C(t)\right| $
is in principle an experimentally observable quantity.\ If the
non-Hermiticity index $\kappa $ is negligible, the two
methods of calculation will give nearly identical result (for typical atoms $%
\kappa $ turns out to be very small, although non-Hermitian issues have
always been ignored from first principles).

The long-range Hamiltonian $H_{0}$ in eq (\ref{1}) contains the
radial Hamiltonian of the colliding electron in a centrifugal
Coulomb potential as well as the free Hamiltonian of the target (an
atomic ion). $f_{i}(\epsilon _{i},r)$ in eq (\ref{4}) is therefore a
Coulomb function regular at the origin (it is also regular at
$+\infty $ only if $\epsilon $ belongs to the spectrum of the radial
part of $H_{0},$ ie when $\epsilon =-1/2n^{2},$ $n\in {N}$).
 The radial channel functions $F_{i}(\epsilon _{i},r)$ appearing in eq (%
\ref{e24}), solutions of the radial part of the redefined $H_0$ for
$r>r_{0}$, are given by a linear combination of Coulomb functions
regular and irregular at the origin, the combination ensuring that
$F_{i}(\epsilon _{i},r)$ converges at $\infty $ \footnote{Note that
$F_{i}(\epsilon _{i},r)$ mathematically diverges as $r\rightarrow
0$, which is of course irrelevant to the scattering problem defined
on $[r_0,+\infty]$}. For the scattering matrix $K(E)$ we take a
$6\times 6$ matrix with a strong energy dependence.\ We also set the
6 values of $\varepsilon _{i}$ to model the internal energies of the
target (we take $\varepsilon _{1}=0$ for the ground state and 5
different values for the excited states of the target). The bound
state energies and coefficients are obtained by applying the
boundary condition $\left\langle r\right. \left| \psi
(E)\right\rangle
\rightarrow 0$ as $r\rightarrow \infty $, yielding the system \cite{seaton}%
\begin{equation}
\left[ K(E)+R(E)\right] Z(E)=0
\end{equation}%
where $R(E)$ is a diagonal matrix with elements $R(E)_{ii}=\tan \pi
(-2(E-\varepsilon _{i}))^{-1}$. This system is solved numerically
for $E$ and then the nontrivial solutions $Z_{i}(E)$ are obtained. \
We compute about $N=400$ eigenstates. The radial overlaps
(\ref{e25}) are determined
analytically, and from there we compute the metric elements $\mathcal{G}%
_{EE^{\prime }}$. For the overall chunk, the non-Hermiticity index is
calculated as $\kappa =0.07$. The ordered distribution of the $N$ largest
off-diagonal elements of the metric is shown in Fig.\ 1.

\begin{figure}[tb]
\begin{center}
\includegraphics[height=2.1in,width=3in]{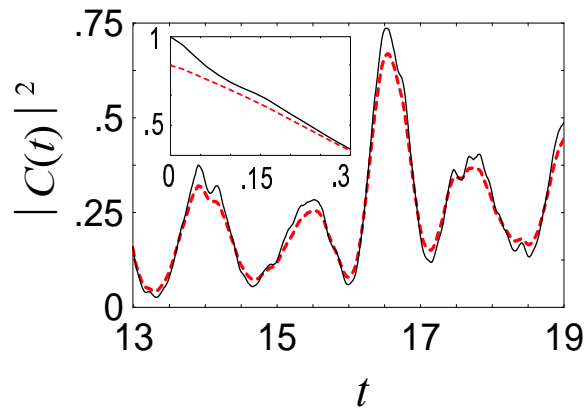}
\caption[]{The autocorrelation function $C(t)$ is given as a
function of time (in units of the period of the classical orbit of
the electron having the mean energy of the initial state). The inset
shows the short-time situation. Dashed (colour online: red) curve:
the autocorrelation function is computed by ignoring the
non-Hermitian character of the problem, following eq (\ref{e67}).
Solid line: Result computed by taking non-Hermiticity into account
(eq (\ref{e70d})). } \label{fig2}
\end{center}
\end{figure}

We now choose an initial state $\left| \zeta (t=0)\right\rangle $,
that we take to be a Gaussian localized radially very far from the
target, at the outer turning point of the radial potential for an
excited electron (with a mean energy $n=55$), with the target being
in its ground state. Initially $\left| \zeta (t=0)\right\rangle$ is
defined on an orthogonal basis of $\mathcal{H}_{ph}$ but we assume
(and verify numerically) that this state
can approximately be expanded on our chunk of computed eigenstates of $%
\widetilde{H}$ as%
\begin{equation}
\left| \zeta (t=0)\right\rangle =\sum a(E)\left| \psi (E)\right\rangle .
\label{e65}
\end{equation}%
where the $a(E)$ are determined by projections. At this point we
proceed along the two different lines mentioned above. In the first
method of calculation we employ the machinery of standard
(Hermitian) quantum mechanics, ignoring non-Hermiticity issues.\
This may appear absurd in view of the preceding discussion, but this
is the way computations are undertaken in applied
problems \footnote{%
It is true that in typical atomic problems, $\kappa$ is
significantly smaller  (below $10^{-3}$) than in the example given
here, so that the computed results would only be marginally affected
by taking into account the non-Hermitian character of the
Hamiltonian.}. Moreover this will allow us to assess the relevance
of the formalism given above in practical calculations -- as we will
see by comparing the first method to the second one, where the
formalism developed in Sec 3 will be employed.

In the first method the expansions
\begin{equation}
\sum_{E}\left| \psi (E)\right\rangle \left\langle \psi (E)\right| \left[ 1%
\textrm{ or }E\textrm{ or }\exp (-iEt/\hbar )\right]  \label{e66}
\end{equation}
are taken as representations of the unit operator, the Hamiltonian
or the evolution operator respectively. The coefficients $a(E)$ of
eq (\ref{e65}) are thus given by the projection of this 'unit'
operator as $\left\langle \psi (E)\right| \left. \zeta
(t=0)\right\rangle $, and the autocorrelation
function (\ref{e63}) follows by employing this 'evolution' operator,%
\begin{equation}
C(t)=\sum_{E}e^{-iEt/\hbar }\left| \left\langle \psi (E)\right|
\left. \zeta (t=0)\right\rangle \right| ^{2}.  \label{e67}
\end{equation}%
The result is shown in Fig.\ 2 by the dashed line; in particular
the inset shows the short-time evolution, and it may be noticed
that at $t=0$ we do not have $C(t=0)=1$, ie $\left| \zeta
(t=0)\right\rangle $ is not normalized after the application of
the 'unit' operator (\ref{e66}), which as we know is not the
correct unit operator on $\mathcal{H}_{ph}$. Neither is the
'evolution' defined by eq (\ref{e66}) unitary: $\left\langle \zeta
(t)\right| \left. \zeta (t)\right\rangle $ computed with eq
(\ref{e66}) shows strong oscillations, displayed in Fig.\ 3.

The correct method to compute $C(t)$ involves first mapping $\left| \zeta
(t=0)\right\rangle $ to $\mathcal{H}$, yielding $\mathcal{G}^{1/2}\left|
\zeta (t=0)\right\rangle $ (cf eq (\ref{e57})), then apply the unitary
evolution operator in $\mathcal{H}$ given by eq (\ref{e50}) and finally
compute the result with the inner product (\ref{e55}) in $\mathcal{H}$. If
we follow the notation (\ref{e57}) and put%
\begin{equation}
\left| \phi (t=0)\right\rangle =\mathcal{G}^{1/2}\left| \zeta
(t=0)\right\rangle  \label{e69}
\end{equation}%
we get the following equivalent expressions for the autocorrelation function:%
\begin{eqnarray}
C(t) &=&\left( \phi (t=0),\widetilde{U}(t)\phi (t=0)\right) _{\mathcal{G}}
\label{e70a} \\
&=&\left\langle \tilde{\phi}(t=0)\right| \widetilde{U}(t)\left| \phi
(t=0)\right\rangle  \label{e70b} \\
&=&\left\langle \phi (t=0)\right| \mathcal{G}^{-1}\widetilde{U}(t)\left|
\phi (t=0)\right\rangle  \label{e70c} \\
&=&\left\langle \zeta (t=0)\right| \mathcal{G}^{-1/2}\widetilde{U}(t)%
\mathcal{G}^{1/2}\left| \zeta (t=0)\right\rangle .  \label{e70d}
\end{eqnarray}%
Eqs (\ref{e70a}) and (\ref{e70b}) give the autocorrelation function as
computed entirely in $\mathcal{H}$ whereas eq (\ref{e70d}) is the same
expression in $\mathcal{H}_{ph}$. $\mathcal{G}^{-1/2}\widetilde{U}(t)%
\mathcal{G}^{1/2}$ appears as the (correct and unitary) evolution operator in $%
\mathcal{H}_{ph}$ resulting from the mapping given by eq
(\ref{e53}). The computed result is shown by the solid line in Fig.\
2, which of course obeys $C(t=0)=1$ (normalization at other times
follows from unitarity).

The most salient feature arising from the comparison of the two curves in
Fig.\ 2 concerns the different profiles of the autocorrelation functions.
This implies that it will not be possible to recover the correct result (\ref%
{e70a}) from the first method result (\ref{e67}) by simply
renormalizing the latter in $\mathcal{H}_{ph}$ (as is sometimes done
in practical scattering problems). Conversely it would not make much
sense to assume that the initial physical state (\ref{e65}) is known
in $\mathcal{H}$, so that one would not need to determine  mapping
$\mathcal{H}\rightarrow \mathcal{H}_{ph}$. Such an exception happens
in  the specific but nevertheless important cases in which one is
only interested in transitions involving given eigenstates of
$\widetilde{H}$.

\begin{figure}[tb]
\begin{center}
\includegraphics[height=2.1in,width=3in]{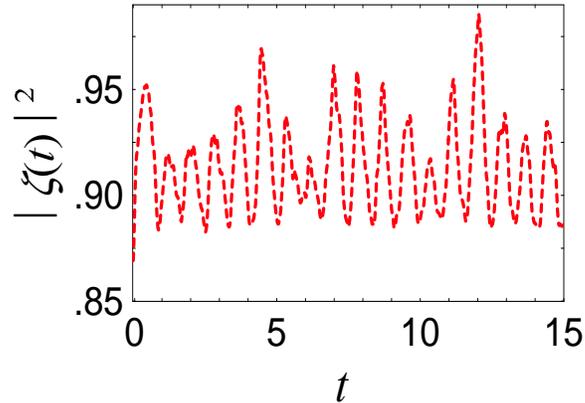}
\caption[]{The 'norm' $\left\langle \zeta (t)\right| \left. \zeta
(t)\right\rangle $ is computed as function of time (in units of the
period of the mean Kepler orbit). The norm is not conserved because
the 'evolution' operator (\ref{e66}) is Hermitian but not unitary in
$\mathcal{H}_{ph}$. } \label{fig3}
\end{center}
\end{figure}

\section{Conclusion}

We have seen that the widely employed formalism of bound state scattering
theory should be properly understood within the framework of non-Hermitian
quantum mechanics. Although in typical cases the non-Hermiticity index $%
\kappa $ is small so that in practice non-Hermitian issues can be
ignored, we have given an illustration for which the calculations of
experimentally observable quantities require the proper
non-Hermitian formulation. The latter has essentially the same
structure and tools as the PT-symmetric systems (reformulated in the
quasi-Hermitian framework) that are currently being extensively
investigated. However in the present case, the physical meaning of
non-Hermiticity is more transparent than in the case of PT-symmetric
quantum-mechanics. In particular, we have seen that changing the
radial interval minutely from $[0,+\infty]$ in the underlying exact
problem to $[r_0,+\infty]$ in the scattering problem leads to an
entire redefinition of the Hilbert spaces relevant for quantum
mechanics. Indeed, by this change the  Hamiltonian $\widetilde{H}$
becomes quasi-Hermitian on $\mathcal{H}_{ph}$. One can then either
redefine the inner product, constructing a new Hilbert space
$\mathcal{H}$, or map the states and operators to the physical
Hilbert space $\mathcal{H}_{ph}$. We have seen that computations are
simpler to undertake in $\mathcal{H}$ than in $\mathcal{H}_{ph}$,
but except in the specific cases involving the sole eigenstates of
the non-Hermitian Hamiltonian, this simplicity is only apparent: as
arbitrary physical states are known in $\mathcal{H}_{ph}$, the
mapping between the two Hilbert spaces must be explicitly determined
anyway, involving the computation of the metric. In the example
given in this work the metric was constructed from the numerical
calculation of the  exact eigenstates of $\widetilde{H}$ in a
restricted energy interval of interest.

The interesting insight gained by the existence of an underlying
exact Hamiltonian $H^{e}$ is that for scattering states, the
physical Hilbert space $\mathcal{H}_{ph}$ is essentially the same as
the Hilbert space of the exact problem. But the expansion in
$\mathcal{H}_{ph}$ of a physical state in terms of the eigenstates
of the \emph{exact} Hamiltonian  differs from the expansion in terms
of the eigenstates $\psi(E)$ of the non-Hermitian Hamiltonian
(although the physical results -- eigenvalues, probability
amplitudes -- will be identical). Actually the expansion in terms of
the eigenstates of the exact Hamiltonian in $\mathcal{H}_{ph}$ is
identical to the expansion in terms of the eigenstates of
$\widetilde{H}$ in $\mathcal{H}$. These remarks suggest that as far
as the scattering eigenstates are concerned, $\mathcal{H}$ is more
physical than $\mathcal{H}_{ph}$, where these eigenstates become
$\mathcal{G}^{-1/2}\left| \psi(E)\right\rangle$.  From a more
general standpoint it appears that quantum mechanics  requires above
all a Hilbert space $\mathcal{H}$ on which the operators are
self-adjoint relative to a given inner product, whatever this inner
product may be. In this work the Hilbert space $\mathcal{H}_{ph}$
defined with the standard inner product (the $L^{2}$ inner product)
only entered the problem because in bound state scattering arbitrary
physical states and operators are already known in this space. In
general however it is possible to envisage the case in which the
standard inner product would not play a special r\^{o}le, although
such a situation will probably lead to intricate interpretational
issues regarding the physical significance of computed quantities.

\end{document}